\documentstyle[psfig,twoside,fleqn,espcrc2]{article}

\newcommand{\AmS}{{\protect\the\textfont2
  A\kern-.1667em\lower.5ex\hbox{M}\kern-.125emS}}

\newcommand{\mrm}[1]{\mathrm{#1}}

\renewcommand{\b}{{\mrm{b}}}
\renewcommand{\c}{{\mrm{c}}}
\renewcommand{\d}{{\mrm{d}}}

\newcommand{\g}{{\mrm{g}}}

\newcommand{\q}{{\mrm{q}}}

\newcommand{\J}{{\mrm{J}}}
\newcommand{\rH}{{\mrm{H}}}
\newcommand{\K}{{\mrm{K}}}

\newcommand{\cbar}{\overline{\mrm{c}}}

\newcommand{\qbar}{\overline{\mrm{q}}}

\newcommand{\mc}{m_{\c}}

\newcommand{\trv}{_{\perp}}
\newcommand{\LQCD}{\Lambda_{\rm{QCD}}}

\begin{document}

\begin{titlepage}

\begin{flushright}
CERN-TH/97-253\\
hep-ph/9709489
\end{flushright}

\vspace{1.5cm}

\begin{center}
\Large\bf Power counting in exclusive quarkonium decays
\end{center}

\vspace{1.2cm}

\begin{center}
Gerhard A.\ Schuler$^a$\\
{\sl Theory Division, CERN, CH-1211 Geneva 23, Switzerland}
\end{center}

\vspace{1.3cm}

\begin{center}
{\bf Abstract}\\[0.3cm]
\parbox{11cm}{
Recent progress in the factorization of exclusive decays of 
heavy quarkonia into long-distance and short-distance contributions 
is presented. The role of higher Fock states and colour-octet 
contributions is outlined.
}
\end{center}

\vspace{1cm}

\begin{center}
{\sl 
To appear in the Proceedings of the\\
High-Energy Physics Euroconference on Quantum Chromodynamics\\
Montpellier, France, 3--9 July 1997}
\end{center}

\vspace{1.5cm}

\vspace{\fill}
\noindent
\rule{60mm}{0.2mm}

\vspace{1mm} \noindent
${}^a$ Heisenberg Fellow.

\vspace{10mm}\noindent
CERN--TH/97--253\\
September 1997


\end{titlepage}

\thispagestyle{empty}
\vbox{}
\newpage

\setcounter{page}{1}


\title{Power counting in exclusive quarkonium decays}

\author{G.A.\ Schuler\address{Theory Division, CERN, CH-1211 Geneva 23, 
Switzerland}\thanks{Heisenberg Fellow.}}
       

\begin{abstract}
Recent progress in the factorization of exclusive decays of 
heavy quarkonia into long-distance and short-distance contributions 
is presented. The role of higher Fock states and colour-octet 
contributions is outlined.
\end{abstract}

\maketitle

\section{INTRODUCTION}

Exclusive reactions at large momentum transfer $Q$ can be calculated 
in perturbative QCD (pQCD) owing to a factorization theorem \cite{Brodsky80} 
that separates the short-distance physics of the partonic subreactions at 
the scale $Q$ from the longer-distance physics associated with the 
binding of the partons inside the hadrons. The full amplitude is given
as a sum of terms, where each term factors into two parts, 
a hard-scattering amplitude $T_{H}$, calculable in 
pQCD, and wave functions $\psi(x_i,\bf{k}_{\perp i})$ for 
each hadron $H$: 
\begin{equation}
  {\cal M} = \sum_{n}\, T^{\rH} \times \prod_i\, \psi_i
\ .
\label{HSAfact}
\end{equation}
The separate dependence on the factorization scale $\mu_F$ 
cancels, to the order considered, in the full amplitude ${\cal M}$. 

The importance of the various terms in (\ref{HSAfact}) depends on their
scaling with $1/Q$. For the contribution that has the weakest fall-off 
with $Q$ (leading-twist contribution), the amplitude $T_{H}$ describes 
the scattering of clusters of {\em collinear} partons from the hadron and 
is given by {\em valence-parton} scatterings only. Hence the only 
non-perturbative input required are the distribution amplitudes (DAs)
$\phi_H(x_i,Q)$ for finding valence quarks in the hadron, each 
carrying some fraction $x_i$ of the hadron's momentum. The DAs
represent wave functions integrated over transverse momentum ${\bf
k}\trv$ up to a factorization scale $\mu_F$, of order $Q$, and obey the 
BL evolution equations \cite{Brodsky80}. Indeed, most calculations so far 
have been performed in this standard hard-scattering approach (sHSA) 
summarized by:
\hfill\\[0.5ex] $\bullet$
leading (i.e.\ valence) Fock states only;
\hfill\\[0.5ex] $\bullet$
collinear approximation; 
\hfill\\[0.5ex] $\bullet$
(mostly) lowest-order in $\alpha_s(Q)$; 
\hfill\\[0.5ex] $\bullet$
tailored for $S$-wave hadrons (orbital angular
 momentum $L=0$).
Up to now results of the sHSA remained at a qualitative level, lacking 
a rigorous discussion of 
(in particular the power-like $\propto 1/Q$) 
corrections. Power corrections can be classified as follows: (i) 
  Corrections arising from the overlap of the soft wave functions;
(ii) 
  corrections associated with the transverse momentum of the partons
  inside the hadrons;
(iii)
  corrections from higher Fock states.

Recently, exclusive processes at large scales have attracted new interest:
\hfill\\[0.5ex] $\bullet$
New, precise data on the pion--photon transition form factor 
allow a quantitative extraction of the (non-perturbative) pion DA 
\cite{JKR96}. 
\hfill\\[0.5ex] $\bullet$
A modified hard-scattering approach (mHSA) 
has been developed, which allows for the incorporation
of transverse degrees of freedom and gluonic radiative corrections in
the form of a Sudakov factor \cite{BoLiSt}. This does not only allow for 
a dynamical setting of the renormalization, $\alpha_s(\mu_{Ri})$,  
and factorization, DA$_i(\mu_{Fi})$, scales. Now also the
perturbative contribution to exclusive observables can be calculated
self-consistently, i.e.\ without major contributions from phase-space 
regions, where the virtualities of the internal partons become 
soft and perturbation theory is not applicable. 
\hfill\\[0.5ex] $\bullet$
A new factorization approach for {\em inclusive} quarkonium decays 
(and production) has been developed \cite{BBL}. Colour-octet contributions 
associated with higher Fock components are found to be crucial. 
\hfill\\[0.5ex] $\bullet$
The importance of higher Fock states also for {\em exclusive} quarkonium 
decays has been realized and new power-counting rules derived \cite{BKS96}. 

\section{{\bf $\chi_{\c J} \rightarrow \pi\pi$} in the mHSA}
\begin{figure}[tb]
\begin{center}
\psfig{figure=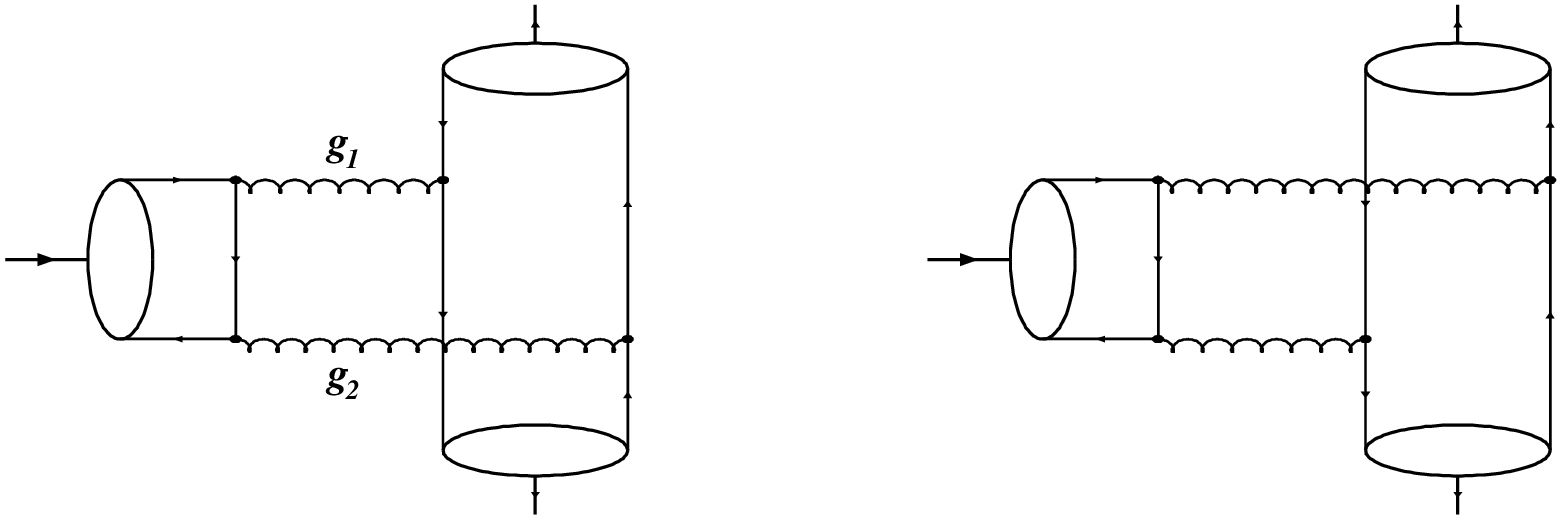,%
       bbllx=50pt,bblly=180pt,bburx=750pt,bbury=240pt,%
       width=0.45\textwidth,clip=} 
\end{center}
 \caption[dummy0]{Feynman graphs for the colour-singlet decay
  $\chi_{\c J} \rightarrow \pi\, \pi$ ($J=0,2$).  
  \label{fig:col1graph} }
\end{figure}

The decay $\chi_{\c J}$ into a pair of pseudoscalar mesons is 
given by the Feynman diagrams in Fig.~\ref{fig:col1graph}, 
describing the annihilation of the dominant Fock state of $\chi_{\c J}$ 
into a pair of gluons. For $J^{PC}=J^{++}$ mesons, the valence Fock state 
is a ${}^3P_J$ colour-singlet $\c\cbar$ pair 
(in the spectroscopic notation ${}^{2S+1}L_J$ of a spin 
angular-momentum state). 
Decays of heavy quarkonia can be calculated in the HSA 
since the heavy-quark mass $\mc$ provides the hard scale $Q = \mc$. 
The sHSA result has been known for a long time \cite{dun80} 
\begin{equation}
\Gamma[\chi_{\c J}\rightarrow \pi\pi] = c_J\, 
\mc\, \alpha_s^4\, \left( \frac{f_\pi}{\mc}\right)^4\, 
 \frac{ |R'_P(0)|^2 }{\mc^5}
\ ,
\label{oldcs}
\end{equation}
where $c_J$ results from a convolution of the pion DAs with the hard 
kernel of Fig.~\ref{fig:col1graph}, $f_\pi$ is the pion decay constant, and 
$R'_P(0)$ the derivative of the coordinate wave function at the origin. 

Recently, the calculations of the decay widths in the mHSA have been 
completed \cite{BKS96}. 
The colour-singlet decay amplitude within the mHSA is 
calculated from the amplitude
\begin{eqnarray}
  M^{(1)}(\chi_{\c J} \to \pi  \pi ) \sim 
 \left(  \frac{\delta_{ij}}{\sqrt{3}}\, 
         \int \d^3 k\, \tilde{\Psi}_J^{(1)} \, S_J^{(1)} \right)
& &\nonumber\\ 
      \int \d x\,  \d y \, \int \d^2 {\bf b}_1\, \d^2 {\bf b}_2
      \exp[-S]
& &\nonumber\\ 
      \hat \Psi_{\pi}^{\ast}(y,{\bf b}_2)\,
      \hat T_{HJ}(x,y,{\bf b}_1,{\bf b}_2)\,
      \hat \Psi_{\pi}(x,{\bf b}_1)\,
\ .
  \label{Mb}
\end{eqnarray}
This is the convolution w.r.t.\ the momentum fractions $x,y$ and 
transverse separation scales ${\bf b}_1,{\bf b}_2$ of the two pions; 
$\hat T_{HJ}$ is the Fourier transform of the hard-scattering amplitude 
$T_{HJ}$ with the transverse ($k\trv$) dependence retained, 
see Fig.~\ref{fig:col1graph}.
The Sudakov factor $\exp[-S]$ takes into account those gluonic radiative
corrections not accounted for in the QCD evolution of the wave
function. Finally, the $\chi_{\c J}$ are treated non-relativistically: 
the bracket in (\ref{Mb}) contains the colour-singlet and spin projection 
operators, and $\tilde{\Psi}_J^{(1)}$ are the (reduced) non-relativistic 
wave functions, which yield 
$R'_P(0)$ after integration over the relative $\c\cbar$ momentum $k$. 
The factorization and renormalization scales depend on the virtualities 
of the respective adjacent fermion propagators, e.g.\ 
\begin{equation}
  \mu^2_{Ri} = \max\left\{4 (1-x) (1-y)\, \mc^2,
     \frac{1}{b^2_1},\frac{1}{b^2_2}\right\} \, .
  \label{tj}
\end{equation}

The results of the improvement of the mHSA calculation over the sHSA one
can be summarized as follows: first, 
(soft) end-point contributions are severely suppressed. 
Indeed, the results are self-consistent in the sense that only about
$2\%$ of the decay rate originate from phase-space regions where 
$\alpha_s(\mu_{R1}) \alpha_s(\mu_{R2}) > 1/2$. Secondly, the uncertainty 
of the result against variation of the input parameters of the calculation 
(shape parameter $B_2$ of the pion-wave function parameter, charm quark mass, 
and QCD scale $\LQCD$) is reduced in the mHSA compared to the sHSA but still
sizeable, see Table~\ref{tab:results}. However, even with an optimistic 
parameter choice the theoretical prediction falls short by a factor of $4$, 
calling for an additional contribution. 
\begin{table}[tb]
\setlength{\tabcolsep}{0.39pc}
\caption[dummy2]{Decay widths of $\chi_{\c J} \to
  \pi^+\pi^-$
  for various choices of the parameters compared with data.}
\label{tab:results}
\begin{tabular}{rrrrr} 
\hline
  $B_2$ & $\mc$ & $\LQCD$ 
  & \multicolumn{2}{c}{$\Gamma(\chi_{\c J} \to \pi^+\pi^-)\,$[keV]  }
\\ \hline
 & [GeV] & [GeV] & $J=0$  & $J=2$  
\\ \hline
  \multicolumn{5}{c}{Modified HSA}
\\ \hline
  $0$ & $1.5$ & $0.2$  
  &   8.22  &  0.41   
\\ 
  $0.1$ & $1.5$ & $0.2$   
  &   12.13  &  0.53 
\\ 
  $-0.1$ & $1.5$ & $0.2$   
  &   5.61  &  0.33  
\\ 
  $0$ & $1.8$ & $0.2$   
  &  2.54  &  0.12 
\\ 
  $0$ & $1.35$ & $0.2$   
  & 15.3   & 0.78  
\\ 
  $0$ & $1.5$ & $0.15$   
  &   4.34  &  0.21
\\ 
  $0$ & $1.5$ & $0.25$   
  &  13.1  &  0.68  
\\ \hline
  \multicolumn{5}{c}{Experiment}
\\   \hline
\multicolumn{3}{c}{PDG\cite{pdg96}}
       &  $105 \pm 30$   &  $3.8 \pm 2.0$   
\\   
\multicolumn{3}{c}{BES\cite{bes96}}
       &  $62.3 \pm 17.3$   &  $3.04 \pm 0.73$   
\\   \hline
\end{tabular}
\end{table} 

\section{Inclusive decays of heavy quarkonia}

The non-relativistic quark-potential model is very successful in describing 
the static properties of heavy quarkonia. A quarkonium $|H(J^{PC})\rangle$ 
is considered to be pure, obviously colour-singlet, 
$|Q\bar{Q}({}^{2S+1}L_J) \rangle$ state where  
the $Q\bar{Q}$-pair is bound by an instantaneous potential. 
The non-relativistic nature of the bound system 
implies the existence of three important scales (besides the heavy-quark 
mass $m$): (the modulus of) the heavy-quark three momentum $\sim mv$, 
the quark's kinetic energy $\sim mv^2$, and the QCD scale $\LQCD$. 

Inclusive quarkonium decays into light hadrons are accessible 
to perturbative QCD: for sufficiently large $m$, the annihilation time 
$\sim 1/m$ is much smaller than the time scales relevant to $Q\bar{Q}$ 
binding. Hence the short-distance annihilation of the $Q\bar{Q}$ pair 
into gluons and light quarks can be separated from the long-distance effects 
of $Q\bar{Q}$ binding. The basis of this separation is provided by an 
effective field theory, NRQCD, supplemented by perturbative factorization 
\cite{BBL}. The inclusive decay rate is given as a double expansion 
in $\alpha_s(m)$ and $v^2$, to leading order in $\LQCD/m$. 

The factorization makes use of the fact that there are, in any Feynman 
diagram, some propagators that are off shell by an amount $\sim m^2$ 
much larger than the scales governing $Q\bar{Q}$ binding. Hence, in the 
effective low-energy theory these may be contracted to a point. The 
remainder of the diagram involves only distances much larger than $1/m$. 
For the case of inclusive decays, these are related to expectation 
values of universal long-distance matrix elements (MEs) ${\cal O}_n$. 
The ${\cal O}_n$ are well-defined 4-fermion operators of NRQCD of the form
${\cal O}_n = (\psi^\dagger K_n \chi) (\chi^\dagger K'_n \psi)$. 
The inclusive decay widths thus take on the form
\begin{equation}
  \Gamma = \sum_n\, f_n\, \langle H | {\cal O}_n | H \rangle
\ ,
\label{Gamincl}
\end{equation}
where the short-distance coefficients $f_n$ are calculable as perturbation 
series in $\alpha_s(m)$.

The importance of the various long-distance MEs in (\ref{Gamincl}) 
is given by their scaling with the heavy-quark velocity $v$. 
The velocity-sca\-ling (po\-wer-counting) rules were first established based 
on consistency requirements of the NRQCD field equations \cite{Lepage92}. 
More rigorously, these follow from a multipole expansion in the wavelength 
of the emitted (multiple) gluon field $\lambda = R_H/\lambda\!\!\!\!{-}_g$. 
In NRQCD, $\lambda \sim k_g/(mv) \sim v$. This formulation has the advantage 
that it can be generalized to the situation where $\LQCD$ is not much
smaller than $mv^2$ (as assumed in NRQCD) giving rise to different 
scaling rules \cite{Newscal}.

The $\J/\psi$ is dominantly an $S$-wave state. Its Fock-state expansion 
starts as
\begin{eqnarray}
  |\J/\psi(1^{--})\rangle & = & O(1)\, |\c\cbar_1({}^3S_1)\rangle
\nonumber\\ & & \hspace*{-30pt}
                      +~ O(v)\, |\c\cbar_8({}^3P_J)\, \g \rangle
                      + O(v^2) 
\ ,
\label{SFockexp}
\end{eqnarray}
where the subscripts $1$ and $8$ denote colour-singlet and colour-octet 
states, respectively. 
Annihilation into light hadrons via the valence 
Fock state involves the ME 
$\langle \J/\psi | {\cal O}_1 (^3S_1) | \J/\psi \rangle$, which we take
to scale as $v^0 = 1$ (dividing out an overall $v^3$ scaling). In the 
potential model this corresponds to $|R_S(0)|^2$. The first corrections 
for $S$-wave quarkonia arise from higher-dimensional operators 
(suppressed by $v^2$) still involving the leading Fock component. These 
lead, for example, to differences between the $\J/\psi$ and $\eta_{\c}$ 
wave functions. The colour-octet component of (\ref{SFockexp}) enters
first at relative order $v^4$. Nonetheless, its contribution can be 
important, namely when the short-distance coefficients of the singlet states
are suppressed by powers of $\alpha_s(m)$ and/or kinematical factors. 
An example is $\J/\psi$ production at high transverse momentum \cite{BFY}. 

The situation is different for $\chi_{\c J}$, as becomes evident 
from their Fock-state expansion
\begin{eqnarray}
  |\chi_{\c J}(J^{++})\rangle & = & O(1)\, |\c\cbar_1({}^3P_J)\rangle
\nonumber\\ & & \hspace*{-30pt}
                      +~ O(v)\, |\c\cbar_8({}^3S_1)\, \g \rangle
                      + O(v^2) 
\ .
\label{Fockexpansion}
\end{eqnarray}
The $v^2$ factor of the $E1$ multipole suppression of the $\c\cbar \g$
state is compensated by the fact that a $P$-wave operator scales 
as $v^2$ relative to an $S$-wave operator. Hence, 
contrary to the case of $S$-wave decays, {\em two} $4$-fermion operators 
contribute to the decay rate of $P$-wave states 
into light hadrons at leading order in $v$
\begin{eqnarray}
 \Gamma[\chi_{\c J} \rightarrow \mathrm{LH}] & = & 
  \frac{c_1}{\mc^4}\, 
     \langle \chi_{\c J} | {\cal O}_1(^3P_J) | \chi_{\c J} \rangle 
\nonumber\\
  & & \hspace*{-50pt} +~ \frac{c_8}{\mc^2}\, 
     \langle \chi_{\c J} | {\cal O}_8(^3S_1) | \chi_{\c J} \rangle 
  + O(v^2 \Gamma)
\ .
\label{eq:inclfact}
\end{eqnarray}
The term involving the colour-singlet matrix element is the one familiar 
from the quark-po\-ten\-tial model. Indeed, up to corrections of order  
$v^2$, one has 
$\langle \chi_{\c J} | {\cal O}_1(^3P_J) | \chi_{\c J} \rangle  = 
9 |R'_P(0)|^2 / (2\pi)$. 

The decays of $P$-wave 
(and higher or\-bi\-tal-an\-gu\-lar-mo\-men\-tum states) 
probe components of the quar\-ko\-ni\-um wave function that involve 
dynamical gluons. However, the contribution to the inclusive 
annihilation rate from the higher Fock state $|\c\cbar \g \rangle$ is 
parametrized through a single number, namely the expectation value 
of the octet operator between the $\chi_{\c J}$ state, i.e.\ 
the colour-octet matrix element in (\ref{eq:inclfact}). 

\section{Factorization of exclusive quarkonium decays}
\begin{figure}[tb]
\begin{center}
\psfig{file=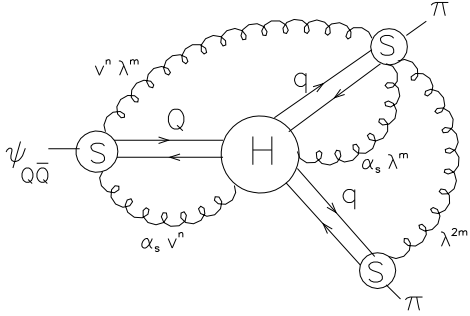,width=0.4\textwidth}
\vspace*{-20pt}
\caption{Factorization of $\chi_{\c J} \rightarrow \pi\pi$.}
\end{center}
\label{fig:exclfact}
\end{figure}
In contrast to inclusive decays, 
there are now two possibilities to absorb long-distance contributions, 
into either quarkonium-specific
or light-hadron-specific objects. 
%
In order to set up a factorization scheme for exclusive quarkonium decays 
one has to consider the virtualities occurring in an arbitrary 
Feynman diagram. Three regions have to be distinguished. 
Propagators can be off shell by order $m^2$, 
$(mv)^2$, or $\LQCD^2 \equiv (\lambda m)^2$. Correspondingly, 
such contributions have to be associated with the hard amplitude, 
the quarkonium, or the light hadrons, respectively. This then
leads to a triple expansion in $\alpha_s(m)$, $v$, and~$\lambda$. 

Another difference with inclusive ha\-dro\-nic decays is the fact that 
the factorization into short-dis\-tance and long-dis\-tance contributions 
holds at the amplitude level rather than at the level of decay rates 
(i.e.\ squared amplitudes). In this sense factorization of exclusive decays
resembles more the case of inclusive electromagnetic decays, where the 
final-state particles are also colour singlets.

\begin{figure}[tb]
\begin{center}
\psfig{figure=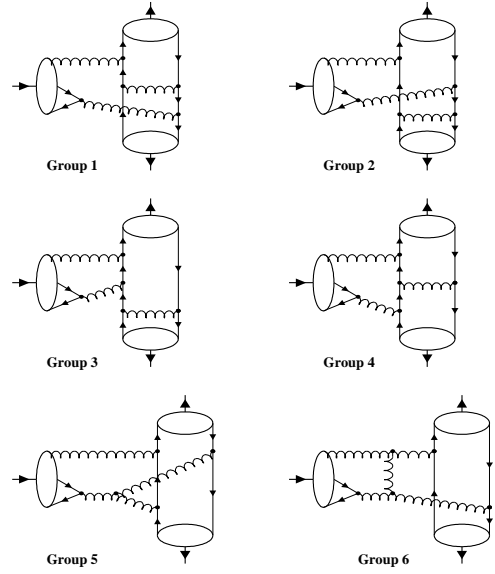,%
       bbllx=18pt,bblly=157pt,bburx=567pt,bbury=797pt,%
       width=0.4\textwidth}
 \caption{Some colour-octet contributions to $\chi_{\c J}\rightarrow \pi\pi$. 
  \label{fig:col8graphs} }
\end{center}
\end{figure}
Higher Fock components in light hadrons certainly exist. 
Consider the decay $\chi_{\c J} \rightarrow \pi\pi$, 
Fig.~\ref{fig:col8graphs}. The (multipole-based) Fock-state expansion 
of the pion starts as follows
\begin{eqnarray}
  | \pi(0^{-+}) \rangle & = & \hspace*{-1.5pt}
  O(1) |\q\qbar_1(^1S_0) \rangle
      + O(\lambda)    |\q\qbar_8(^1P_1)\g \rangle
\nonumber\\ & & \hspace*{-50pt} 
  +~ O(\lambda^2)\, \left\{ |\q\qbar_8(^3S_1)\g \rangle
  + \ldots \right\} + O(\lambda^3)
\ .
\label{Fockpion}
\end{eqnarray}
Since both octet terms give corrections to the leading one 
of the same order $\lambda^2$,  
two new (non-perturbative) light-cone DAs had
to be specified in a calculation using higher Fock components of the pion. 
Clearly, not much predictive power can be expected. In fact, 
higher Fock contributions to the pion\footnote{
The situation may be different for decays into higher-$L$ mesons.}
are power-suppressed ($\propto \LQCD^2/m^2$) and hence become negligible 
for sufficiently large quark masses. 

In contrast, higher Fock components of heavy quarkonia are suppressed 
by powers of $v$ only, or even scale as the leading Fock contribution. 
Moreover, owing to the non-relativistic expansion one does not need 
new DAs but only new `decay constants' (expectation values of NRQCD 
operators). Of course, there remains the problem of colour conservation. 
A perturbative factorization can be set up for the case
$\alpha_s(\mu)/\pi \ll 1$ for $\mu \sim mv$ since this allows one to 
treat the constituent gluon(s) as the constituent (valence) quarks. 
Colour conservation is then achieved 
by coupling the constituent gluon (in all possible ways) to the hard 
process, see Fig.~\ref{fig:col8graphs}. 

The calculation of $\chi_{\c J} \rightarrow \pi\pi$ in the collinear 
approximation leads, however, to singular results. This is not surprising 
as the Feynman diagrams of Fig.~\ref{fig:col8graphs} must contain 
contributions, which, to leading order in $\alpha_s$, constitute the 
two higher Fock states $\q\qbar\g$ of the pion. The regularization of the
propagators associated with the `would-be' constituent gluon of the pion 
depends on the hierarchy of $\lambda$ and $v^2$: the singularities are 
protected by either relativistic correction $\propto mv^2$ to the 
heavy-quark momentum or the finite transverse momentum $k_{\perp} 
\propto \LQCD = \lambda m$ of the (light) valence quarks of the pion. 

\begin{table}[tb]
\setlength{\tabcolsep}{0.39pc}
 \begin{center}
 \caption[]{Results for the $\chi_{\c J}$ decay widths (in keV) into pions
            ($f^{(8)} = 1.46 \times 10^{-3} \,\mbox{GeV}^2 \,$; $B_2=0$)
            in comparison with experimental data. 
}
\label{tab:widths}
  \begin{tabular}{lccc} 
\hline
      &\phantom{results}  &  PDG \cite{pdg96} & BES \cite{bes96} 
\\ \hline
  $\chi_{\c 0} \to \pi^+ \pi^-$  & 45.4  
     & $105 \pm 47\phantom{6}$ & $64 \pm 21\phantom{.3}$ 
\\ 
  $\chi_{\c 2} \to \pi^+ \pi^-$  & 3.64
     & $3.8 \pm 2.0$ & $3.04 \pm 0.73$ 
\\ 
  $\chi_{\c 0} \to \pi^0\,\pi^0$  &  23.5
     & $43 \pm 18$ & 
\\ 
  $\chi_{\c 2} \to \pi^0\,\pi^0$ &  1.93
     & $2.2 \pm  0.6$  & 
\\ \hline
  \end{tabular}
 \end{center} 
\end{table}
In \cite{BKS96} $v^2 \ll \lambda$ ($\ll v$) was assumed, in which case 
it is the non-zero value of $k_{\perp}$ taken into account in the mHSA 
that acts as a regulator and, in turn, determines the size of the 
colour-octet contribution to the decay width. The only new non-perturbative 
parameter is then a single number, namely the octet decay constant $f^{(8)}$
entering the $|\c\cbar\g\rangle$ wave function 
\begin{equation}
  |\chi_{\c J}^{(8)} \rangle 
\hspace*{-0.5pt} =  \hspace*{-0.5pt}
 \frac{t^a_{ij}}{2} f^{(8)} 
   \hspace*{-0.5pt}   \int  \hspace*{-0.5pt}
   \d z_1\d z_2 \Phi_J^{(8)}(z_1,z_2,z_3)  S_{J\nu}^{(8)}
\label{coc}
\end{equation}
where $t=\lambda/2$ is the Gell-Mann colour matrix and $a$ the colour
of the gluon. 
With this single parameter $\chi_{\c J}$ decays into all pairs of 
pseudoscalar mesons can be predicted. The $f^{(8)}$ value obtained by 
fitting to the four decays into pions (Table~\ref{tab:widths}) 
corresponds to a very reasonable probability of the octet 
$|\c\cbar\g\rangle$ state, $P_{\c\cbar\g} \sim 0.5$. 
Predictions for $\chi_{\c J}$ decays into $\K$'s and $\eta$'s also  
agree nicely with data, see Table~\ref{tab:kaon}.
\begin{table}[tb]
 \begin{center}
 \caption[]{Results for the $\chi_{\c J}$ decay widths into $\K$'s
            and $\eta$'s in comparison with experimental data
            ($f^{(8)} = 1.46 \times 10^{-3} \,\mbox{GeV}^2 \;$, $B_1^\K=0$).}
\label{tab:kaon}
\begin{tabular}{lcc} 
\hline
      & $J=0$ & $J=2$ 
\\ \hline 
    & \multicolumn{2}{c}{$\Gamma[\chi_{\c J} \to \K^+ \K^-]\;\;$ [keV]}  
 \\ \hline
    $B_2^\K = -0.176$   & 22.4  & 1.68  
\\ 
    $B_2^\K = -0.100$   & 38.6  & 2.89  
\\ \hline
    PDG \cite{pdg96} & $99 \pm 49$  & $3.0 \pm 2.2$ 
\\ 
    BES \cite{bes96} & $52 \pm 17$  & $1.04 \pm 0.43$
\\ \hline
    & \multicolumn{2}{c}{$\Gamma[\chi_{\c J} \to \eta \eta]\;\;$ [keV]} 
\\ \hline
    $B_2^{\eta}\; = -0.036$   & 24.0 & 1.91  
\\ 
    $B_2^{\eta}\; = \phantom{-}0.000$   & 32.7 & 2.66  
\\ \hline
    PDG \cite{pdg96} & $35 \pm 20$  & $1.6 \pm 1.0$ 
\\ \hline
  \end{tabular}
 \end{center} 
\end{table}
Predictions for $\Upsilon$ decays have been obtained as well 
(Table~\ref{tab:Ydecay}). 

It should be emphasized that the colour-octet contribution 
from the $|\c\cbar\g\rangle$ Fock state is numerically important, 
in accordance with the fact that it is not suppressed relative to  
the contribution from the valence Fock state, 
either in $\LQCD/m$ or in $v$. 
\begin{table}[tb]
\begin{center}
\caption{Decay widths in eV for $P$-wave bottomonia for two values 
of $f^{(8)}$ ($m_b=4.5\,$GeV;  $R'_P(0)=0.7\,$GeV${}^{5/2}$).}
\label{tab:Ydecay}
\begin{tabular}{lcc}
\hline
 $f^{(8)}\, [\, {\mrm{GeV}}^2\, ]$ &
 $\chi_{\b 0} \rightarrow \pi^+\pi^-$ &
 $\chi_{\b 2} \rightarrow \pi^+\pi^-$ 
\\ \hline
  $1.46 \times 10^{-3}$ & $20.6$ & $1.69$
\\ 
  $5 \times 10^{-3}$ & $152$ & $13.8$
\\ \hline
  $f^{(8)}\, [\, {\mrm{GeV}}^2\, ]$ &
 $\chi_{\b 0} \rightarrow \pi^0\pi^0$ &
 $\chi_{\b 2} \rightarrow \pi^0\pi^0$ 
\\ \hline
  $1.46 \times 10^{-3}$  & $10.5$ & $0.88$
\\ 
  $5 \times 10^{-3}$  & $78.2$ & $7.27$
\\ \hline
\end{tabular}
\end{center}
\end{table}

\section{Summary}
It is now possible to obtain accurate predictions for exclusive 
reactions involving pseudoscalar mesons. On the one hand, 
the non-perturbative input, their DAs, can well be determined 
from recent precise measurements of the meson--photon transition 
form factors. On the other hand, with mHSA an improved framework 
exists for calculations of the contributions of the leading Fock state. 
Hence, the leading colour-singlet contributions are well under control.

In the case of the charmonium $\chi_{\c J}$ decays into a pair of 
pseudoscalars, one finds that the colour-singlet contribution alone 
is not sufficient to accommodate the data. Indeed, it turns out that 
the colour-octet contribution from the $|\c\cbar\g\rangle$ Fock state 
contributes at the same level as the colour-singlet one. Its inclusion 
yields good agreement with experimental data. 

In general, one can expect colour-octet contributions to be important 
if the contribution from the leading Fock state is suppressed. 
This happens for quarkonia with orbital-angular momentum $L \geq 1$. 
It occurs also if decays of the leading Fock state into massless quarks 
($m_{q} = 0$) are forbidden by a symmetry (such as $G$-parity or helicity) 
so that these seemingly leading decays are actually power-suppressed 
$\propto (m_{q}/m_{Q})^n$. 

The study \cite{BKS96} is a first attempt to derive a factorization 
approach for exclusive decays of heavy quarkonia. More work is needed 
to generalize the factorization to higher orders in $\alpha_s(m_{Q})$, 
$v$, and $\LQCD/m$. In particular it has to be shown that the various 
contributions in an arbitrary Feynman diagram can be disentangled 
in a systematic way.\hfill\\[0.5ex]

\noindent
{\it Acknowledgements:}
\hfill\\ 
I am pleased to thank J.\ Bolz and P.\ Kroll for an enjoyable collaboration.

\end{document}